\documentclass[fleqn,12pt,twoside]{article}
\usepackage[headings]{espcrc1}
\usepackage{amssymb}

\readRCS
$Id: espcrc1.tex,v 1.2 2004/02/24 11:22:11 spepping Exp $
\usepackage{graphicx}
\usepackage[figuresright]{rotating}

\newcommand{\AmS}{{\protect\the\textfont2
  A\kern-.1667em\lower.5ex\hbox{M}\kern-.125emS}}

\title{RHIC results from LHC perspectives}

\author{Carlos A. Salgado\address[MCSD]{Department of Physics, CERN,
Theory Division\\ CH-1211 Geneva, Switzerland}}
       
\runtitle{RHIC results from LHC perspectives} 
\runauthor{Carlos A. Salgado}

\begin{document}

\maketitle

\begin{abstract}
RHIC data opens new ways of characterizing the medium created in a heavy
ion collision by measuring particles with high transverse momentum. In
addition to the observation of the predicted jet quenching, the new data on
particle correlations indicate that a strong modification of the jet-like 
shapes is driven by the dynamical properties of the medium. I will review the 
lessons we are learning from RHIC and outline some new directions 
which could become of primary importance at the LHC.
\end{abstract}

\section{Introduction}

Quarks or gluons produced at high transverse momentum in elementary collisions
build up their hadronic wave function by emitting partons, mostly gluons, 
reducing their large virtuality ($Q\sim p_t$) to a
typical hadronic scale. These emitted particles form QCD parton showers which
can be experimentally observed as jets. The
properties of these objects can be computed by resumming the relevant logarithms
originated by the large difference in the scales of the problem. A well-known
example of these resummations are the DGLAP evolution equations describing the
virtuality dependence of the fragmentation functions (for a general
introduction on these topics see e.g. \cite{Ellis:1991qj}).
In Monte Carlo event generators the virtuality
evolution of the created particles is a major issue and several solutions are
proposed to match these parton showers with known results at a fixed
order in perturbation theory. Jets being spatially extended objects (or
equivalently, sensitive to different scales) provide optimal conditions 
for the characterization of the medium created in a nucleus-nucleus collisions.
The main predictions for high-$p_t$ measurements in a nuclear environment
are the suppression of the inclusive particle yields due to the additional
medium energy-loss and the broadening of the associated jets
\cite{Baier:1996sk,Zakharov:1997uu,Wiedemann:2000za,Gyulassy:2000er,Wang:2001if,reviews}.
 Indeed, the strong
suppression of high-$p_t$ particles \cite{Adcox:2001jp}
is one of the main observations made at 
RHIC which now extends to particles up to $p_t\sim$ 20 GeV \cite{shimom}. 
The situation for the jet
broadening is, however, not as clear, due basically to the limited range of
virtualities studied in present experiments. The future LHC will largely
improve this limitation
 thanks to a logarithmically extended range of virtualities.

On the theory side, there is a wide consensus that the dominant mechanism of
{\it jet quenching} is the medium-induced gluon radiation
\cite{Baier:1996sk,Zakharov:1997uu,Wiedemann:2000za,Gyulassy:2000er,Wang:2001if,reviews}. 
The way the inclusive gluon spectrum can be translated into a 
medium-modified parton shower
has not been fully solved theoretically and in most cases the independent
gluon emission approximation leading to a Poisson distribution is employed.
This formalism reproduces successfully  
\cite{reviews,d'Enterria:2005cs} 
the observed suppression of hadrons with $p_t\sim 5...20$ GeV.
The main outcome of these analyses is meant to be
the density of the produced medium. In the following we will briefly introduce
the formalism, its main successes and weakness and comment on how present
experimental data can be described and the new possibilities open at both RHIC 
and the LHC.


\section{The formalism}


{\it The vacuum}.
Parton branching is an essential property of QCD processes involving large
virtualities. It is present in the initial or final state radiation leading to 
the scale evolution of the parton distributions or fragmentation functions
respectively and is well reproduced by DGLAP evolution equations. 
An equivalent integral formulation for the evolution
with the virtuality $t$ of these quantities
\begin{equation}
f(x,t)=\Delta(t)f(x,t_0)+\int^t_{t_0}\frac{dt'}{t'}\frac{\Delta(t)}{\Delta(t')}
\int\frac{dz}{z}\frac{\alpha_S}{2\pi}P(z)f(x/z,t'),
\label{eqevol}
\end{equation}
provides a simple probabilistic 
interpretation in terms of the non-branching probabilities 
({\it Sudakov form factors})
\begin{equation}
\Delta(t)\equiv\exp\left [-\int_{t_0}^t\frac{dt'}{t'}\int dz
\frac{\alpha_S}{2\pi}P(z)\right],
\label{eqsudak}
\end{equation}
where $P(z)$ are the QCD splitting functions. These equations
resum the relevant divergences of the vacuum
radiation spectrum $\omega dI^{\rm vac}
/d\omega dk_t^2\sim 1/k_t^2$ of gluons with
energy $\omega$ and transverse momentum $k_t$, being $z=\omega/E$ and 
$E$ the energy of the parent parton.


{\it The medium.}
Eqs. (\ref{eqevol}) and (\ref{eqsudak}) describe the process in which a parton
reduces its energy and virtuality by radiating other
partons (gluons in general) with a fraction of momentum $z$. The description
of the experimental data given by 
this general formalism is one of the most precise
tests of QCD. The counterpart for the branching of a parton traversing 
a medium has not yet 
reach the same degree of refinement.
In the following we will restrict to the
case of final state radiation. In most current approximations,
one simply neglects the
virtuality ordering of the process, resumming the energy loss of
subsequent branchings by a simple Poisson distribution 
\cite{Baier:2001yt,Salgado:2003gb,Gyulassy:2001nm}
\begin{equation}
  P_E(\epsilon) = \sum_{n=0}^\infty \frac{1}{n!}
  \left[ \prod_{i=1}^n \int d\omega_i \frac{dI^{\rm med}(\omega_i)}{d\omega}
    \right]
    \delta\left(\epsilon -\sum_{i=1}^n {\omega_i \over E} \right)
    \exp\left[ - \int d\omega \frac{dI^{\rm med}}{d\omega}\right].
  \label{eqqw}
\end{equation}
These {\it quenching weights} \cite{Baier:2001yt,Salgado:2003gb}
give the probability of an additional
energy loss, $\Delta E=\epsilon\,E$, for a known medium-induced gluon
radiation spectrum 
 $\omega dI^{\rm med}/d\omega$
\cite{Baier:1996sk,Zakharov:1997uu,Wiedemann:2000za,Gyulassy:2000er,Salgado:2003gb}.
 Assuming that the hadronization
takes place outside the medium, the medium-modified
fragmentation functions are computed by the convolution of (\ref{eqqw}) with the
vacuum fragmentation functions 
\cite{Wang:1996yh,Gyulassy:2001nm,Salgado:2002cd}. Neglecting the evolution is
a reasonable approximation for moderate values of $p_t$ in which the
correction due to energy loss is expected to dominate over
the milder (logarithmic) evolution in virtuality \cite{Salgado:2002cd}.

\subsection{The medium-induced gluon radiation spectrum}

The general description of a high-energy parton propagation
is in terms of Wilson lines \cite{Hebecker:1999ej,Wiedemann:2000za,reviews}, 
averaged in the allowed configurations of a
medium. Several prescriptions for these medium averages exist, but most of them
are based in a model of the medium as a
collection of static scattering centers with no correlations
between them \cite{Gyulassy:1993hr}. 
In this framework, at high-energy, the effect of the each individual scattering
reduces to induce color rotations and to transfer 
transverse momentum.
For large number of scatterings ({\it multiple soft
scattering}) a saddle point approximation of the scattering amplitudes 
defines the single parameter of the model, the transport coefficient 
$\hat q$, with the meaning of the average
transverse momentum transferred to the gluon per mean free path.

The spectra of gluons computed in this framework can be found in 
\cite{Baier:1996sk,Zakharov:1997uu,Wiedemann:2000za,Gyulassy:2000er,Wang:2001if,reviews,Salgado:2003gb}
An interesting property
is that, contrary to the vacuum case, the medium-induced spectrum
is infrared and collinear finite due to formation time effects
\cite{Salgado:2003gb,Salgado:2003rv}. The functional
dependences in $\omega$ and $k_t$ away from the regions affected by formation
time are, however, stronger than the vacuum ones. In particular the $k_t$ 
dependence is that
of a typical elastic scattering $dI^{\rm med}/dk_t^2\sim 1/k_t^4$, 
-- in other words, this
corrections appear as higher-twist -- and, also, the energy spectrum
of radiated gluons is softer than in the vacuum.

The angular dependence of the radiation is affected by formation time effects
in a peculiar way. Recalling that $t_{\rm form}\sim \omega/k_t^2$ 
the typical transverse momentum for a static medium
\begin{equation}
k_t^2\simeq\int^{t_{\rm form}} d\xi \hat q(\xi)=\hat q t_{\rm form}\ \ 
\Longrightarrow\ \ k_t^2\sim \sqrt{\omega\hat q}
\end{equation}
defines a typical emission angle
$
\sin\theta\sim\left(\frac{\hat q}{\omega^3}\right)^{1/4}
$
. Radiation is then suppressed for $\omega\lesssim \hat q^{1/3}$
\cite{Salgado:2003gb}.
For a very opaque medium (large $\hat q$) the typical radiation angle is large
for small $\omega$ and reaches its maximum at the end of the phase space for
$\omega\lesssim \hat q^{1/3}$. These estimations are independent on the length
of traversed medium\footnote{Alternatively to the multiple soft scattering
approximation, an
expansion in scattering centers is often used 
\cite{Gyulassy:2000er,Wang:2001if}.
In this case the emission angle decreases proportional to the
length of traversed medium, implying more collinear radiation for larger
in-medium pathlength.}

\section{Inclusive particle suppression}

Due to the different coupling to the emitted gluon, the quark energy loss is
known to be smaller by a factor $C_F/N_C=4/9$ than the gluon energy loss.
Massive quarks are also known to lose
less energy due to the {\it dead cone effect} \footnote{This name
refers, originally, to
the suppression of collinear emission in the vacuum due to mass terms. 
In the medium, the larger massive quark formation time
leads to a {\it filling} of the already suppressed collinear region 
\cite{Armesto:2003jh}. However, 
this small effect don't affect the discussion here.}
\cite{Dokshitzer:2001zm,Armesto:2003jh,Zhang:2003wk,Djordjevic:2003zk}. 
This translates into a quantified identity
dependence of the nuclear modification factor $R_{AA}$, defined as the
suppression with respect to the scaling with the number of collisions.

\begin{figure}
\vskip -1.5cm
\begin{center}
\includegraphics[width=0.5\textwidth,angle=-90]{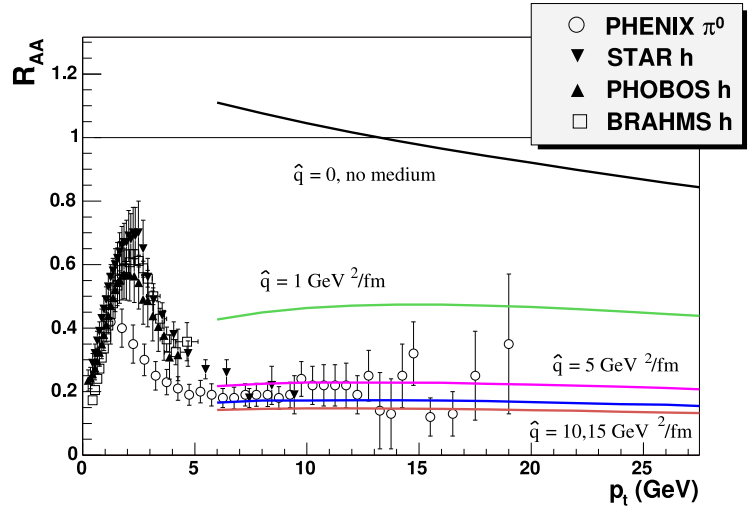}
\end{center}
\vskip -0.5cm
\caption{Nuclear modification factor $R_{AA}$ for charged hadrons in central
AuAu collisions at $\sqrt{s}$=200 GeV \protect\cite{Eskola:2004cr}.
All data from Ref. \cite{Adcox:2001jp} except preliminary PHENIX data 
from \cite{shimom}.} 
\label{figincl}
\vskip -0.3cm
\end{figure}

Experimental data on light-meson suppression for central AuAu collisions at
RHIC are used in Fig.~\ref{figincl} to find the best value of the time
averaged transport coefficient \cite{Eskola:2004cr,Dainese:2004te}. 
However, when the values of $\hat q$ exceed 
$\sim$5 GeV$^2$/fm, $R_{AA}$ is dominated by geometry and the
sensitivity to the actual value of $\hat q$ is reduced
(see Fig. \ref{figincl}). This
surface emission \cite{Eskola:2004cr,Dainese:2004te,Muller:2002fa}
is originated by the steeply falling perturbative partonic 
spectrum which due to triggering in high-$p_t$ particles select 
those which lose small amounts of energy.
This is an intrinsic limitation of inclusive particle
production for a precise measurement of the energy loss.


\begin{figure}
\begin{center}
\includegraphics[width=0.4\textwidth,angle=90]{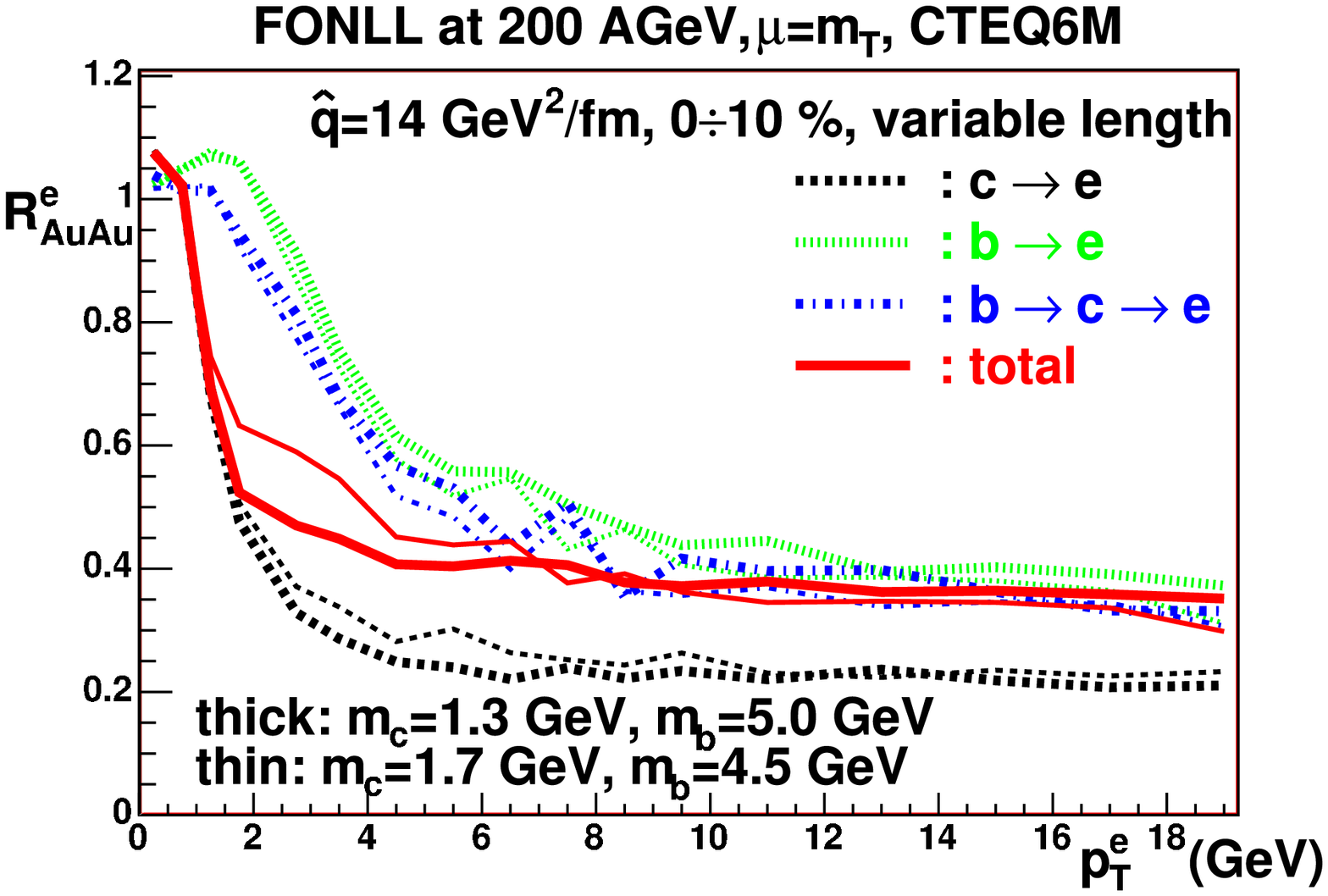}
\caption{$R_{AA}$ for electrons coming from the decay of charm quarks (black
line) beauty (blue and green lines) and total (red) assuming the relative
normalization of $b$ and $c$ as given by the FONLL approximation. The bands
for each curve correspond to the uncertainty on the heavy quark masses.}
\end{center}
\label{figHQ}
\vskip -0.3cm
\end{figure}

One way of improving the determination of the transport coefficient is by
changing the identity of the parent parton, in particular via heavy quarks.
Experimentally, a measurement of identified heavy
mesons in nucleus-nucleus collisions has not been possible yet. A way out is
to detect the weak decays of these mesons into electrons, whose spectrum is,
at RHIC energies, expected to be the dominant contribution, once the photonic
background is subtracted. In Fig. \ref{figHQ} the suppression factor for the 
electrons coming from decays of charm and beauty is computed \cite{noshq} 
using the value of $\hat q$ obtained to fit the $R_{AA}$ for light mesons.
The total contribution of 
$c+b\,\to\,(e^-+e^+)$ is also computed in the FONLL approximation 
\cite{Cacciari:2001td}. Interestingly the mass effects are found to be small
for the charm quark and their suppression almost identical to that of light
quarks \cite{Armesto:2005iq}. Preliminary data presented at this
conference \cite{hqqmstar,hqqmphen} 
find a strong reduction of the non-photonic electron yields, 
compatible with that from the 
decays of charm quarks in Fig.~\ref{figHQ}. This could
indicate a enlarged contribution of the charm spectrum as compared with the
theoretical expectations given by FONLL \cite{Cacciari:2005rk}, but probably
compatible with proton-proton measurements \cite{pphq}. Notice, that in this
case, the mass dependence of the
medium-induced gluon radiation cannot be access with this measurement and the
identification of beauty mesons will be needed \cite{Armesto:2005iq}.
These theoretical uncertainties (see also \cite{Djordjevic:2005db}),
as well as the experimental discrepancies reported at the
conference \cite{hqqmstar,hqqmphen} must be solved before a firm conclusion 
can be drawn. 

\section{Jets in heavy ion collisions?}

The potential of high-$p_t$ measurements as a tool to characterize the medium
properties is not reduced to inclusive particles. The additional induced
radiation modifies the structure of the branching process -- which carries the
signature of the medium -- and can be measured as in-medium
modification of jet properties. A main issue in jet reconstruction (even
in more elementary collisions) is the energy calibration. The amount of
energy deposited into a jet cone of radius
$R\equiv\sqrt{\Delta\phi^2+\Delta\eta^2}$ is
affected by two opposite effects,  namely the 
out-of-cone fluctuations (which increase with the cone radius) and the 
background fluctuations (which decrease with the cone radius).
At RHIC an energy background of 20 GeV (50 GeV)
is expected for jet radius $R$=0.3
(0.5) respectively for central AuAu collisions. The corresponding figures
for the LHC are estimated to be
$E^{\rm bg}\sim$ 100 GeV (250 GeV). A proper subtraction of
these large backgrounds remains the main challenge of jet measurements in heavy
ion collisions. At the LHC \cite{gustaf}, 
the different collaborations ALICE, CMS and ATLAS
study different methods of background subtraction using moderate jet radius
$R=$ 0.3...0.5. From a theoretical point of view, it is, then, 
essential to identify
jet observables in which the medium modification is not largely affected by
the background. 

\begin{figure}
\begin{center}
\includegraphics[width=0.4\textwidth]{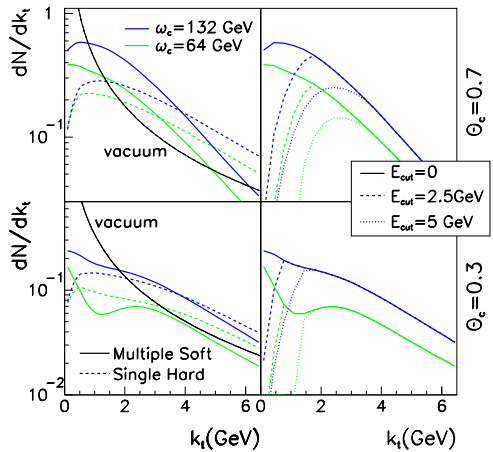}
\caption{The gluon multiplicity distribution inside a cone size $R=\Theta_c$
as a function of the transverse momentum $k_t$ with respect to the jet axis.
In the two right panels the sensitivity to infrared cuts is shown to
be small.}
\vskip -0.3cm
\end{center}
\label{figjet}
\end{figure}

In \cite{Salgado:2003rv} 
the first study of this type of observables was performed. In
particular a small broadening in the energy distributions inside a 
jet was found
when computing the fraction $\rho(R)$ of the total jet energy deposited 
within a subcone of radius $R$. Physically, this means that
the gluons emitted at larger angles
are softer and unable to redistribute a sizable amount of energy. As a
consequence,
the additional out-of-cone fluctuations are not enhanced dramatically.
 This general result is
independent on imposing small-$p_t$ cuts to the observed associated radiation. 
If the jet energy distribution is not 
modified by medium-effects the different structure of the radiation should
manifest in the multiplicity distributions. Indeed, a large broadening is seen
in Fig. \ref{figjet} for the
additional number of medium-induced gluons as a function of their transverse
momentum with respect to the jet axis.
By imposing different cuts to the energy spectrum the sensitivity of this
observable to background subtraction is shown to be small.
 It is worth noting that even though the present calculations
lack of several physical mechanisms as hadronization etc, 
the main conclusions are
independent on the actual realization of the model and are based solely on the
general properties of the medium-induced gluon radiation; in particular the
non-divergency of the spectrum in the infrared and collinear limits. These
properties are given by formation time effects and kinematics.

\subsection{Particle correlations: RHIC}

At RHIC, where a complete jet reconstruction has not been possible, jet-like
structures are measured by two-particle correlations. In this situation, 
the background subtraction is a complicated issue which 
makes the interpretation of the
experimental data difficult in some cases. One clear step forward 
is the first measurement, presented in this conference,
of correlations of particles with $p_t$
above any significant background, namely $8<p_t^{\rm trig}<15$ GeV and
$p_t^{\rm assoc}>$ 6 GeV \cite{magest}. 
The results of this analysis indicates that the near-side peak is
not suppressed, the away side is suppressed as much as $R_{AA}$ and there is
no evidence of broadening in any of them. Moreover, the distributions in
$p_t^{\rm assoc}/p_t^{\rm trig}$ do not present any centrality dependence or
difference with the ones in proton-proton. In the most natural interpretation,
the observed particles come from vacuum fragmentation of parent 
partons
which suffered small medium energy-loss. Two competing trigger-bias effects
lead to this picture: on the one hand, the higher-$p_t$ trigger particle is
produced close to the surface, as explained previously, due to the large slope
of the perturbative spectrum; on the other hand, for the same reason, 
the opposite jet is only observed in the case it suffers small energy loss,
i.e. for small in-medium pathlengths. As a result, only those pairs produced
close and {\it tangential} to the surface can be observed 
\cite{Muller:2002fa,loiz}. 
Calculations
including the quenching weights defined in (\ref{eqqw}) with realist geometry
confirm this picture \cite{loiz}. From the point of view of the formalism, this
measurement is able to constrain the probability of no-interaction in
(\ref{eqqw}). Moreover,
a small energy loss must show up as moderate energy ($\omega\lesssim \hat
q^{1/3}\sim {\cal O}$(1 GeV)) particles correlated with the pair. It would be
interesting to measure this associated radiation by three-particle
correlations provided the background subtraction is experimentally under
control.

{\it Removing the small-$p_t$ cut-off}. At RHIC a series of measurements try
to reconstruct the spectrum of associated particles with small or moderate
$p_t$.
The most spectacular findings are in the away-side jet with the possibility 
of a two-peak structure \cite{Adler:2005ee,Adams:2005ph}. 
This associated radiation was
also found to be softer than in the vacuum \cite{Adams:2005ph}. 
The interpretation of these data
is still not clear, and different mechanisms have been recently proposed 
\cite{conical} which
couple dynamical medium properties to the jet structure. Let us simply mention
here that  these findings do not disagree with qualitative expectations from
the medium-induced gluon radiation which, as explained, predicts
softer spectra of associated particles and large angle emission for gluons 
with $p_t\sim\omega\lesssim \hat q^{1/3}$. A complete implementation of
these effects is still missing; this leads us to our last subject:

\section{Limitations and future developments}

The formalism described in previous sections provide a nice description
of experimental data. It presents, however, limitations related with both its
range of applicability and the potential to study important dynamical 
properties of the medium. I will present here two lines
of future research which, from my point of view, could improve the
present limitations and be of primary importance for the LHC. 

\subsection{Improving the model of the medium}

\begin{figure}
\begin{minipage}{0.52\textwidth}
\includegraphics[width=\textwidth]{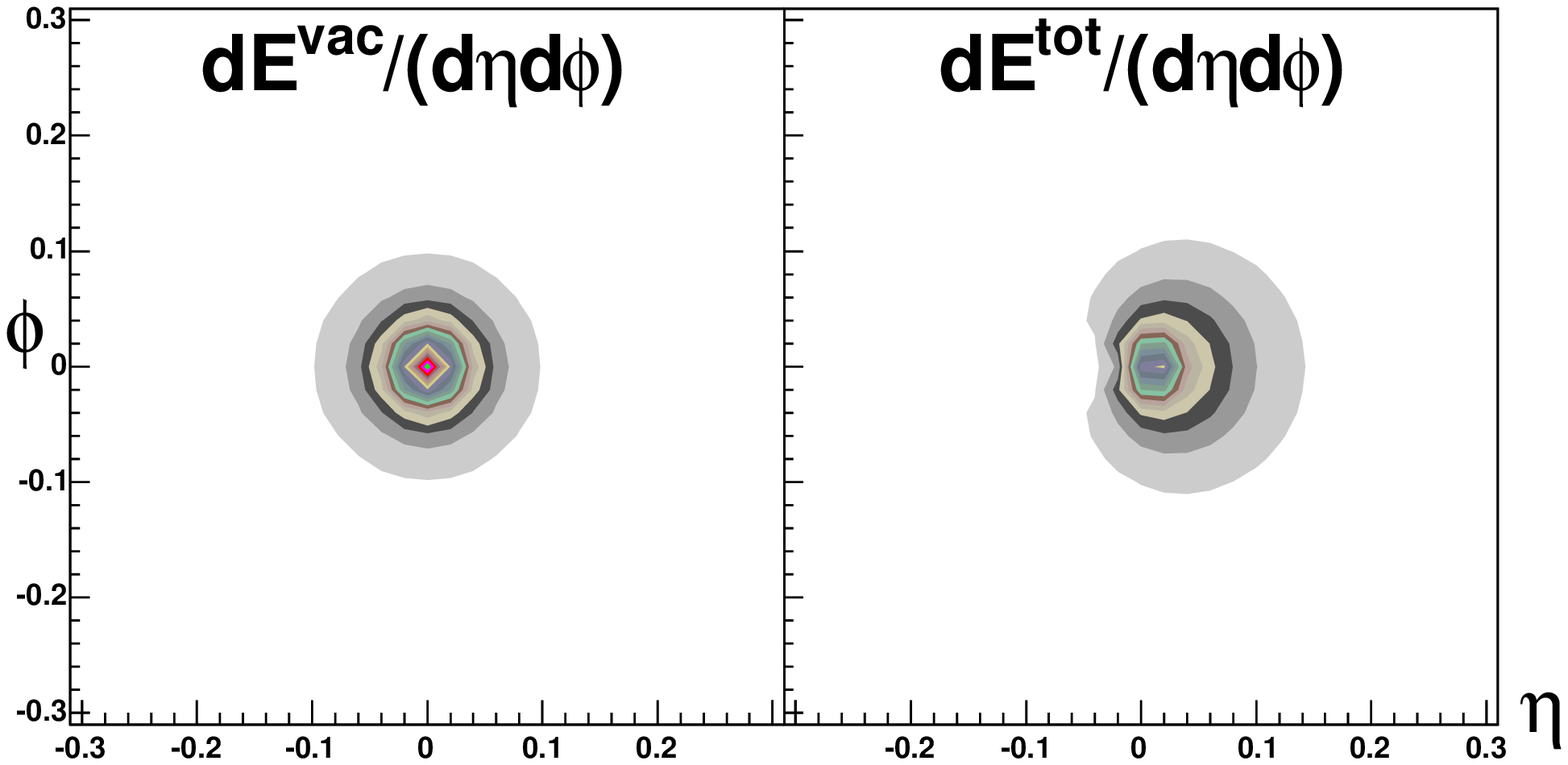}
\caption{Jet energy distributions in the $\eta\times\phi$ plane for a 100 GeV
jet for a non-flowing (left panel) and flowing (right panel) medium.}
\label{figflow}
\end{minipage}
\hfill
\begin{minipage}{0.46\textwidth}
\includegraphics[width=\textwidth]{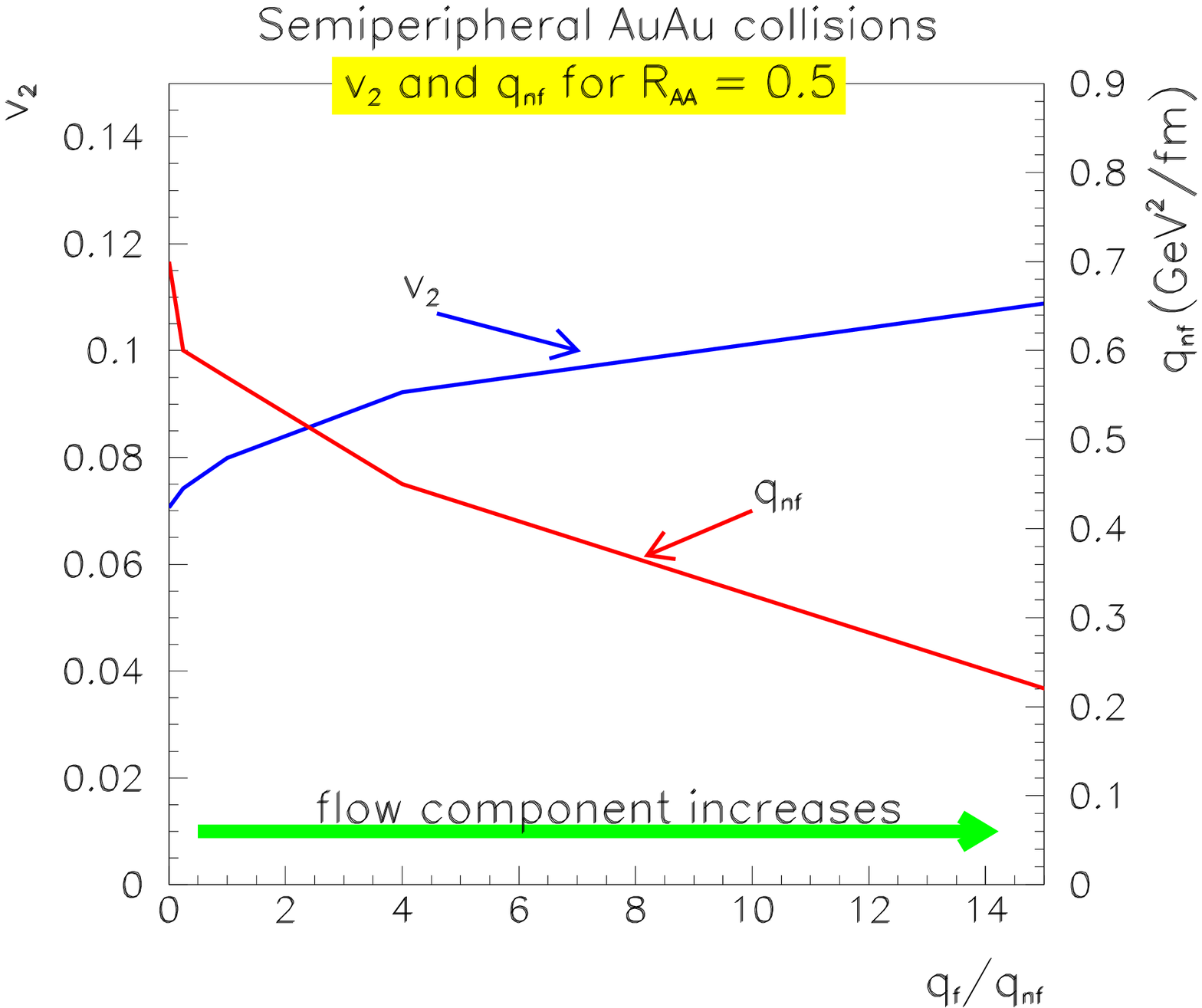}
\caption{Dependence of $v_2$ and non-flow component for the same 
$R_{AA}=0.5$ in semiperipheral AuAu collisions.}
\label{figv2}
\end{minipage}
\vskip -0.3cm
\end{figure}

Hydrodynamical models \cite{hydro}
have revealed to be in nice agreement with experimental
data from RHIC at small transverse momenta. They provide a consistent
picture of the medium created in nucleus-nucleus collisions by means of the
energy-momentum tensor 
\begin{equation}
T^{\mu\nu}=\left(\epsilon+p\right)u^\mu u^\nu-p g^{\mu\nu}
\end{equation}
and an equation of state $p=p(\epsilon,t,\mu_B)$. If $T^{\mu\nu}$ is the
fundamental object which describes the medium, it must determine $\hat q$.
Theoretically, this relation should appear in the
medium-averages of the Wilson lines
describing the propagation of the high energy particles in the medium. In an
exploratory study \cite{Armesto:2004pt}
we have proposed a modification of the scattering potential
to take into account flow-induced gluon radiation. One observable
consequence is the presence of
asymmetric jet-shapes in the $\eta\times\phi$ plane with an elongation in the
direction of the local flow (see Fig. \ref{figflow}). Preliminary data from
the STAR collaboration shows an enhancement and elongation in the
$\eta$-direction of the particle multiplicities associated to a high-$p_t$
trigger \cite{etamag}
in qualitative agreement with the presence of a strong coupling of the
medium-induced gluon radiation to the longitudinal flow. Similar effects are
expected due to the transverse flow. To estimate them we 
generalize \cite{Armesto:2004pt}
the relation \cite{Baier:2002tc} $\hat q=c\epsilon^{3/4}(p)$ -- where
$\epsilon(p)$ is the equation of state -- to $\hat q=c\epsilon^{3/4}(T^{n_\perp
n_\perp})$, which can be written as a sum of a (symmetric)
non-flow plus a (asymmetric) flow components
$\hat q=\hat q_{nf}+\hat q_f({\bf r})$. The presence of the flow component
produces an additional energy loss, translating into additional suppression
of inclusive yields, which affects the estimation of the medium densities
(non-flow component) as shown in Fig.~\ref{figv2}. The asymmetric dependence of
$\hat q_f$ on the azimuthal angle enhances moderately the value of $v_2$ as
well.

\subsection{Improving the shower evolution}

In order to go beyond the independent gluon emission approximation given by
the Poisson distribution (\ref{eqqw}) both energy constrains and the
possibility of secondary branchings must be included. In the vacuum, the shower
evolution as given by Sudakov form factors (\ref{eqsudak}) provide a good
description of the experimental data. For the medium, only a couple of attempts
in this direction are available. On one hand, in a formal derivation 
\cite{Wang:2001if} the medium effects are computed as higher-twist terms 
in a generalization of the DGLAP equations for the nuclear fragmentation
functions in DIS. In this approach modified splitting functions are obtained
which encode the main properties of the radiation as LPM
suppression, etc. The generalization of this formalism to describe other
jet properties has not been worked out yet. On the other hand, in an attempt 
to describe the associated radiation, an enhancement by a constant factor of the
singular part of the vacuum splitting functions is proposed in 
\cite{Borghini:2005em}. In this 
simple approach a good description of $R_{AA}$ is obtained together with a
softening of the associated multiplicity spectrum.

\section{Final comments}

{\it The value of the transport coefficient}.
The large value of $\hat q$ obtained by fits to the experimental data reported
in previous sections deserves some interpretation. In \cite{Baier:2002tc} the
transport coefficient for an ideal QGP gas was computed to be $\hat
q=c\,\epsilon^{3/4}$ with $c_{QGP}^{\rm ideal}\simeq 2$. Assuming a Bjorken
expanding medium the time averaged transport coefficient $\hat q=$ 5...15
GeV$^2$/fm translates into a proportionality factor $c\, >5\, c_{QGP}^{\rm
ideal}$ \cite{Eskola:2004cr}. 
This disagreement opens two interesting possibilities:
 recalling that 
$\hat q$ is not only proportional to the density of scattering
centers but also to the individual scattering cross sections, the first 
possibility is that
the propagating particles couple to the medium with cross
sections at least five times larger than the pertubative ones -- this
interpretation goes in the line of strong non-perturbative effects in the
medium \cite{shur,Molnar:2001ux}; a second possibility is that the presence
of flow fields, which contribute
to the energy loss as explained in previous section,
affects the above estimates. Following this idea,
in an recent work \cite{Renk:2005ta} 
the proportionality
factor $c\sim 2$ is recovered when 
 the flow field contribution to $\hat q$ as proposed in \cite{Armesto:2004pt}
is included.

In summary, RHIC has shown the potential of the high-$p_t$ particles to 
characterize the medium properties with a quantified probe. The dialectic between experiment and
theory opens a large window for new developments in which the
interplay between the soft bulk and the hard part of the spectra will be 
exploited.
We have outlined some of the main directions 
in which the formalism should be improved for a more complete characterization
of the dynamical medium created in high energy heavy ion collisions. The
qualitative new regimes of large virtuality accessible at the LHC will be 
an ideal situation for these studies.

\vskip 0.2cm
{\bf Acknowledgments:} Discussions with N. Armesto, A. Dainese, P.
Jacobs, C. Loizides and U. Wiedemann were a great help for the preparation of
this talk and manuscript.


\begin{thebibliography}{9}

\bibitem{Ellis:1991qj}
  R.~K.~Ellis, W.~J.~Stirling and B.~R.~Webber,
  Camb.\ Monogr.\ Part.\ Phys.\ Nucl.\ Phys.\ Cosmol.\  {\bf 8} (1996) 1.

%
\bibitem{Baier:1996sk}
R.~Baier, Y.~L.~Dokshitzer, A.~H.~Mueller, S.~Peigne and D.~Schiff,
Nucl.\ Phys.\ B {\bf 484} (1997) 265.
%
\bibitem{Zakharov:1997uu}
B.~G.~Zakharov, 
JETP Lett.\  {\bf 65} (1997) 615.
\bibitem{Wiedemann:2000za}
U.~A.~Wiedemann,
Nucl.\ Phys.\ B {\bf 588} (2000) 303.
%
\bibitem{Gyulassy:2000er}
M.~Gyulassy, P.~Levai and I.~Vitev,
Nucl.\ Phys.\ B {\bf 594} (2001) 371.
%
\bibitem{Wang:2001if}
X.~N.~Wang and X.~f.~Guo,
Nucl.\ Phys.\ A {\bf 696} (2001) 788.
%

\bibitem{reviews} For recent reviews see:
  R.~Baier, D.~Schiff and B.~G.~Zakharov,
  Ann.\ Rev.\ Nucl.\ Part.\ Sci.\  {\bf 50}, 37 (2000);
%
  M.~Gyulassy, I.~Vitev, X.~N.~Wang and B.~W.~Zhang,
  arXiv:nucl-th/0302077;
%
  A.~Kovner and U.~A.~Wiedemann,
  arXiv:hep-ph/0304151;
%
  C.~A.~Salgado,
  Mod.\ Phys.\ Lett.\ A {\bf 19} (2004) 271;
  P.~Jacobs and X.~N.~Wang,
  Prog.\ Part.\ Nucl.\ Phys.\  {\bf 54} (2005) 443

                                                                                
\bibitem{Adcox:2001jp}
K.~Adcox {\it et al.}  [PHENIX Collaboration],
Phys.\ Rev.\ Lett.\  {\bf 88} (2002) 022301;
%
S.~S.~Adler {\it et al.}  [PHENIX Collaboration],
Phys.\ Rev.\ C {\bf 69} (2004) 034910;
%
C.~Adler {\it et al.} [STAR Collaboration],
Phys.\ Rev.\ Lett.\  {\bf 89} (2002) 202301;
%
J.~Adams {\it et al.}  [STAR Collaboration],
Phys.\ Rev.\ Lett.\  {\bf 91} (2003) 172302;
%
B.~B.~Back {\it et al.}  [PHOBOS Collaboration],
Phys.\ Lett.\ B {\bf 578} (2004) 297;
%
I.~Arsene {\it et al.}  [BRAHMS Collaboration],
Phys.\ Rev.\ Lett.\  {\bf 91} (2003) 072305.

\bibitem{shimom} M. Shimomura [PHENIX Collaboration] these proceedings.

\bibitem{d'Enterria:2005cs}
  D.~d'Enterria,
  Eur.\ Phys.\ J.\ C\ {\bf 43} (2005) 295

\bibitem{Baier:2001yt}
  R.~Baier, Y.~L.~Dokshitzer, A.~H.~Mueller and D.~Schiff,
  JHEP {\bf 0109} (2001) 033

\bibitem{Salgado:2003gb}
  C.~A.~Salgado and U.~A.~Wiedemann,
  Phys.\ Rev.\ D {\bf 68} (2003) 014008 

\bibitem{Gyulassy:2001nm}
  M.~Gyulassy, P.~Levai and I.~Vitev,
  Phys.\ Lett.\ B {\bf 538} (2002) 282

\bibitem{Wang:1996yh}
  X.~N.~Wang, Z.~Huang and I.~Sarcevic,
  Phys.\ Rev.\ Lett.\  {\bf 77} (1996) 231

\bibitem{Salgado:2002cd}
  C.~A.~Salgado and U.~A.~Wiedemann,
  Phys.\ Rev.\ Lett.\  {\bf 89} (2002) 092303 

\bibitem{Hebecker:1999ej}
  See e.g. A.~Hebecker,
  Phys.\ Rept.\  {\bf 331} (2000) 1 and references therein.

\bibitem{Gyulassy:1993hr}
  M.~Gyulassy and X.~n.~Wang,
  Nucl.\ Phys.\ B {\bf 420} (1994) 583

\bibitem{Salgado:2003rv}
  C.~A.~Salgado and U.~A.~Wiedemann,
  Phys.\ Rev.\ Lett.\  {\bf 93} (2004) 042301 

\bibitem{Dokshitzer:2001zm}
Y.~L.~Dokshitzer and D.~E.~Kharzeev,
Phys.\ Lett.\ B {\bf 519}  (2001) 199.
                                                                                
\bibitem{Armesto:2003jh}
N.~Armesto, C.~A.~Salgado and U.~A.~Wiedemann,
Phys. Rev. D {\bf 69} (2004) 114003.
                                                                                
\bibitem{Zhang:2003wk}
B.~W.~Zhang, E.~Wang and X.~N.~Wang,
Phys.\ Rev.\ Lett.\  {\bf 93} (2004) 072301.
                                                                                
\bibitem{Djordjevic:2003zk}
M.~Djordjevic and M.~Gyulassy,
Nucl.\ Phys.\ A {\bf 733} (2004) 265.
                                                                                
\bibitem{Eskola:2004cr}
  K.~J.~Eskola, H.~Honkanen, C.~A.~Salgado and U.~A.~Wiedemann,
  Nucl.\ Phys.\ A {\bf 747} (2005) 511 

\bibitem{Dainese:2004te}
  A.~Dainese, C.~Loizides and G.~Paic,
  Eur.\ Phys.\ J.\ C {\bf 38} (2005) 461

\bibitem{Muller:2002fa}
  B.~Muller,
  Phys.\ Rev.\ C {\bf 67}, 061901 (2003)

\bibitem{hqqmstar}
STAR Collaboration: J. Belcik these proceedings; H.-B. Zhang, these 
proceedings.

\bibitem{hqqmphen}
S.~A.~Butsyk [PHENIX Collaboration] these proceedings.

\bibitem{noshq}
N.~Armesto, A.~Dainese, M.~Cacciari, C.~A.~Salgado and U.~A.~Wiedemann, these 
proceedings and work in preparation.

\bibitem{Cacciari:2001td}
  M.~Cacciari, S.~Frixione and P.~Nason,
  JHEP {\bf 0103} (2001) 006

\bibitem{Armesto:2005iq}
  N.~Armesto, A.~Dainese, C.~A.~Salgado and U.~A.~Wiedemann,
  Phys.\ Rev.\ D {\bf 71} (2005) 054027

\bibitem{Djordjevic:2005db}
  M.~Djordjevic, M.~Gyulassy, R.~Vogt and S.~Wicks,
  arXiv:nucl-th/0507019; M.~Djordjevic, these proceedings.

\bibitem{Cacciari:2005rk}
  M.~Cacciari, P.~Nason and R.~Vogt,
  arXiv:hep-ph/0502203.

\bibitem{pphq} Y. Kwon [PHENIX Collaboration] these proceedings.

\bibitem{gustaf} H.~A.~Gustafsson, these proceedings.

\bibitem{magest} D. Magestro [STAR Collaboration] these proceedings; 
P. Jacobs and M. van  Leeuwen [STAR Collaboration] these proceedings;
J. Dunlop, these proceedings.

\bibitem{loiz} A. Dainese, C. Loizides, G. Paic, Acta 
Phys. Hung. A, Heavy Ion Phys. to appear.

\bibitem{Adler:2005ee}
  S.~S.~Adler {\it et al.}  [PHENIX Collaboration],
  arXiv:nucl-ex/0507004;
H. B\"usching, these proceedings; N. Grau, these proceedings; N.~N.~Ajitanand,
these proceedings.

\bibitem{Adams:2005ph}
  J.~Adams {\it et al.}  [STAR Collaboration],
  arXiv:nucl-ex/0501016;
  F.q. Wang, these proceedings; J. Ulery, these proceedings.



\bibitem{conical}
 H.~Stoecker,
  Nucl.\ Phys.\ A {\bf 750}, 121 (2005);
J.~Casalderrey-Solana, E.~V.~Shuryak and D.~Teaney,
  arXiv:hep-ph/0411315
and these proceedings;
J. Ruppert and B.~Muller,
  arXiv:hep-ph/0503158
and these proceedings; 
  V.~Koch, A.~Majumder and X.~N.~Wang,
  arXiv:nucl-th/0507063
and these proceedings;
  I.~M.~Dremin, L.~I.~Sarycheva and K.~Y.~Teplov,
  arXiv:nucl-ex/0509002;
 and these proceedings.

\bibitem{hydro} Y. Hama, these proceedings.

\bibitem{Armesto:2004pt}
  N.~Armesto, C.~A.~Salgado and U.~A.~Wiedemann,
  Phys.\ Rev.\ Lett.\  {\bf 93} (2004) 242301;
  arXiv:hep-ph/0411341.

\bibitem{etamag}
D. Magestro, proceedings of the Hard Probes 2004 Conference, Ericeira 
(Portugal) 2004; P. Jacobs, Eur.\ Phys.\ J.\ C {\bf 43} (2005) 467.

\bibitem{Baier:2002tc}
  R.~Baier,
  Nucl.\ Phys.\ A {\bf 715} (2003) 209


\bibitem{Borghini:2005em}
N.~Borghini and U.~A.~Wiedemann,
arXiv:hep-ph/0506218 and these proceedings.

\bibitem{shur}
E.~V.~Shuryak, these proceedings.

\bibitem{Molnar:2001ux}
  D.~Molnar and M.~Gyulassy,
  Nucl.\ Phys.\ A {\bf 697} (2002) 495
  [ibid. A {\bf 703} (2002) 893]

\bibitem{Renk:2005ta}
  T.~Renk and J.~Ruppert,
  arXiv:hep-ph/0507075.



\end{thebibliography}
\end{document}